\documentclass[aps,twocolumn,showpacs]{revtex4}
\usepackage{amssymb,amsmath}
\usepackage{bm}
\usepackage{graphicx}
\begin{document}

\title{Effect of structural defects on anomalous ultrasound
propagation in solids during second-order phase transitions}

\author{
\firstname{Pavel~V.}~\surname{Prudnikov},
\firstname{Vladmir~V.}~\surname{Prudnikov},
\firstname{Evgenii~A.}~\surname{Nosikhin} }
\affiliation{%
Omsk state university, Department of theoretical physics, pr.Mira 55A, Omsk 644077, Russia
}%

\email{prudnikp@univer.omsk.su}

\date{\today}

\begin{abstract}
The effect of structural defects on the critical ultrasound
attenuation and ultrasound velocity dispersion in Ising-like
three-dimensional systems is studied. A field-theoretical
description of the dynamic effects of acoustic-wave propagation in
solids during phase transitions is performed with allowance for both
fluctuation and relaxation attenuation mechanisms. The temperature
and frequency dependences of the scaling functions of the
attenuation coefficient and the ultrasound velocity dispersion are
calculated in a two-loop approximation for pure and structurally
disordered systems, and their asymptotic behavior in hydrodynamic
and critical regions is separated. As compared to a pure system, the
presence of structural defects in it is shown to cause a stronger
increase in the sound attenuation coefficient and the sound velocity
dispersion even in the hydrodynamic region as the critical
temperature is reached. As compared to pure analogs, structurally
disordered systems should exhibit stronger temperature and frequency
dependences of the acoustic characteristics in the critical region.
\end{abstract}


\maketitle

\section{Introduction}

The progress in understanding the nature of critical phenomena is
mainly related to the theoretical and experimental studies of
critical dynamics in condensed matter. However, the descriptions of
the nonequilibrium behavior of systems during phase transitions
still contain a number of unsolved problems. This is due to the fact
that studying the dynamic properties of critical fluctuations, which
have anomalously high amplitudes and slow damping, encounters
problems that are more complex than the problems that arise when
equilibrium properties are described. Qualitatively, this is caused
by the necessity of taking into account the interactions between
order-parameter fluctuations and other longlived excitations.

The dynamics of phase transitions contains a number of physically
important processes that are determined by the behavior of a
multispin correlation function and, thus, are particularly complex
for a theoretical description. These are, for example, thermal
processes near the critical point in a liquid–gas system, the
attenuation of electromagnetic-field energy that accompanies
magnetic resonance phenomena, and the anomalous attenuation and
scattering of acoustic waves in various media during phase
transitions. The latter processes are important, since they underlie
resonance and ultrasonic methods of studying critical dynamics.

\begin{figure*}
\includegraphics[width=0.95\textwidth]{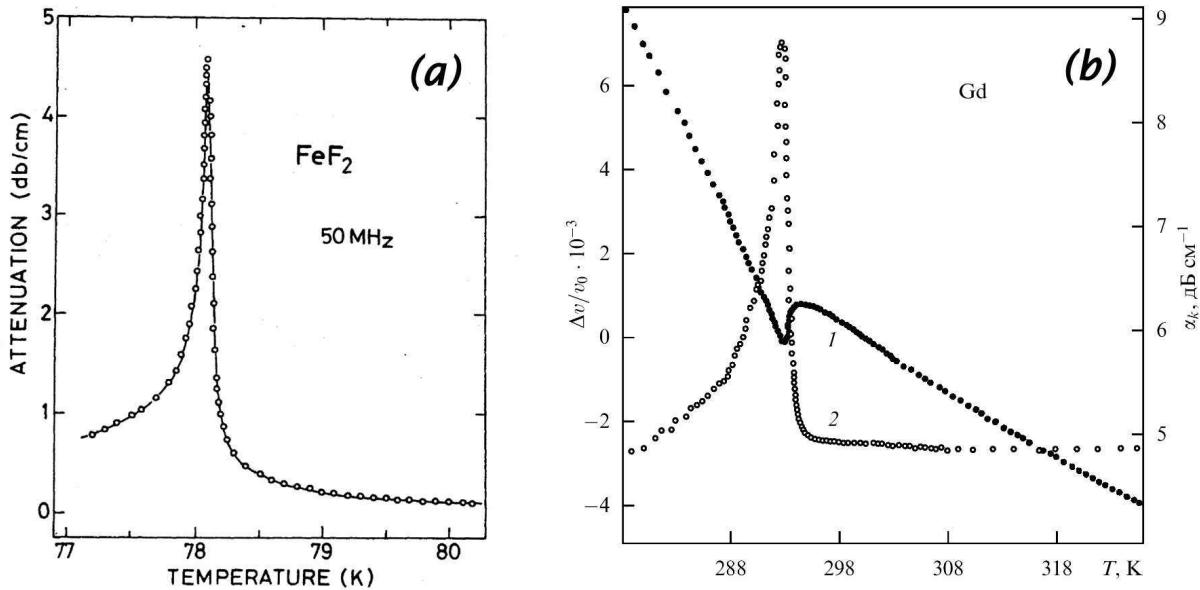}
\caption{ \label{fig:1} Results of the experimental studies of (a)
the ultrasound attenuation coefficient in $\mathrm{\mathop{FeF_2}}$
\cite{IkushimaF} and (b) the ultrasound attenuation coefficient and
ultrasound dispersion in $\mathrm{\mathop{Gd}}$ \cite{Aliev}. }
\end{figure*}

The unique feature of ultrasonic methods is the fact that, at
temperatures that are close to a second-order phase transition
temperature in magnetic systems and systems with structural phase
transitions, researchers detect both anomalously strong ultrasound
attenuation and an anomalous change in the ultrasound velocity,
which can easily be observed in experiment (Figs.~\ref{fig:1}a,
\ref{fig:1}b) \cite{IkushimaF,Aliev}. These phenomena are caused by
the interaction of low-frequency acoustic oscillations with
longlived order-parameter fluctuations, which produce a random force
that disturbs normal acoustic regimes by means of magnetostrictive
spin–phonon interaction. In this process, relaxation and fluctuation
attenuation mechanisms can be distinguished. The relaxation
mechanism, which is due to a linear dynamic relationship between
sound waves and an order parameter \cite{LandauKh}, manifests itself
only in an ordered phase, where the statistical average of the order
parameter is nonzero. Since the relaxation of the order parameter
near a phase-transition point proceeds slowly, this mechanism plays
an important role in the dissipation of low-frequency acoustic
oscillations. The fluctuation attenuation mechanism, which is
determined by a quadratic relation between the deformation variables
in the Hamiltonian of a system with order-parameter fluctuations,
manifests itself over the entire critical-temperature range. To
date, there exist a considerable number of works that deal with a
theoretical description of the ultrasonic anomalies that appear in
condensed matter during phase transitions
\cite{Pawlak,Schwabl93,Kamilov98} and give an adequate explanation
of experimental results \cite{Bhatt,Luthi,Suzuki82}.

One of the most interesting and important problems from both
experimental and theoretical viewpoints is the study of the
influence of structural defects on the ultrasound propagation
characteristics in materials undergoing phase transformations. The
structural disorder induced by impurities or other structural
defects plays a key role in the behavior of real materials and
physical systems. The matter of particular interest is the effect of
frozen structural defects, whose presence can manifest itself in a
random perturbation of a local transition temperature, as it occurs,
for instance, in ferro- and antiferromagnetic systems in the absence
of an external magnetic field. The statistical features inherent in
systems with a frozen disorder create considerable difficulties for
both the analytical description of the behavior of such systems and
the experimental methods of their investigation. According to the
heuristic Harris criterion \cite{Harris74}, the effect of frozen
point defects becomes noticeable and induces a new type of critical
behavior if the critical exponent for the heat capacity of a pure
system is positive. As was demonstrated in previous studies, this
criterion is only met for Ising-like systems. Therefore, the effect
of point structural defects on the critical behavior is negligible
in systems with a multicomponent order parameter, such as the XY
model and the Heisenberg model. Therefore, one of the most
challenging problems from a physical point of view is the study of
the effect of structural defects on the critical behavior of systems
with a single-component order parameter, in which the presence of a
structural disorder leads to a substantial change in the
critical-behavior characteristics.

However, the problem of the effect of structural defects on the
characteristics of ultrasound propagation in materials that undergo
phase transformations still remains unsolved because of the
complexity of the theoretical description of the four-spin
correlations of order-parameter fluctuations, which determine
acoustic characteristics. Pawlak and Fechner \cite{PawlakFecher89}
attempted to describe the effect of point defects on the ultrasound
propagation parameters near a critical temperature using the
first-order $\varepsilon$ expansion. However, as was shown in our
work \cite{PrudnikovCM}, some mistakes crept in the description of
this phenomenon in \cite{PawlakFecher89}; in particular, they used
wrong diagrams for taking into account the dynamic effects of the
interaction of order-parameter fluctuations via a defect-induced
field and they did not use diagrams that give a noticeable
contribution to the attenuation coefficient. Moreover, earlier
investigations \cite{Prudnikov01_03,HighOrder} based on the
field-theoretical description of pure and disordered systems in
two-loop and higher approximations with the use of the method of
summation of asymptotic series demonstrated that the results
obtained in the lowest-order $\varepsilon$ expansion can only be
considered as a crude estimate, especially for disordered systems.
Thus, the results obtained in \cite{PawlakFecher89} require
reevaluation using a more precise approach. For this purpose, in
this work we performed a correct field-theoretical description of
the effect of structural defects on the anomalous critical
ultrasound attenuation and the anomalous change in the ultrasound
velocity in three-dimensional Ising-like compressible systems with
allowance for both the fluctuation \cite{PrudnikovCM} and relaxation
attenuation mechanisms without using the $\varepsilon$-expansion
method.

\section{The model}

For phase transitions in compressible systems, the relation between
an order parameter and elastic deformations is an important factor.
As was first shown in \cite{Larkin69}, the critical behavior of
compressible systems with a quadratic striction is unstable in
regard to the relation between an order parameter and acoustic
modes, and a first-order phase transition that is close to a
second-order phase transition is realized. However, as was clarified
in \cite{Ymry74}, the conclusions made in \cite{Larkin69} are valid
only at low pressures and, beginning with a certain threshold
pressure, the deformation effects induced by an external pressure
change the order of the phase transition.

The Hamiltonian of a disordered compressible Ising model can be
written as
\begin{equation} \label{ham:1}
H = H_{el} + H_{op} + H_{int} + H_{imp}.
\end{equation}
The contribution of the deformation degrees of freedom
is determined as
\begin{multline}
H_{el}=\frac{1}{2}\int {\rm d^{d}}x\, \left(C^{0}_{11}\sum\limits_{\alpha} u^{2}_{\alpha \alpha} \right. +\\
  +\,2C^{0}_{12}\sum\limits_{\alpha\beta} u_{\alpha \alpha} u_{\beta \beta}
  + \left. 4C^{0}_{44} \sum\limits_{\alpha <\beta}u^{2}_{\alpha\beta}\right),
\end{multline}
where $u_{\alpha \beta}(x)$ are the components of the deformation
tensor and $C_{ij}^{k}$ are the elastic constants. The use of an
isotropy approximation for $H_{el}$ is caused by the fact
that, in the critical region, the system behavior parameters
are determined by an isotropic fixed point of renormalization-
group transformations, while the anisotropy
effects are negligible \cite{Izym}.
The magnetic component $H_{op}$ is represented in the form of the Ginzburg–Landau–Wilson Hamiltonian
\begin{equation}
H_{op} = \int  {\rm d^{d}}x\, \left[ \frac{1}{2}\tau_0 S^{2} + \frac{1}{2}\left(\nabla S\right)^{2} +
\frac{1}{4}u_0 S^{4} \right],
\end{equation}
where $S(x)$ is the spin order parameter,
$u_0$ is a positive interaction constant, and
$\tau_0 = {(T - T_{0c})}/{T_{0c}}$ is the
reduced phase-transition temperature. The $H_{int}$ component
determines spin-phonon interaction,
\begin{equation}
H_{int} = \int {\rm d^{d}}x\, \left[ g_{0} \sum_{ \alpha} u_{\alpha \alpha} S^{2} \right],
\end{equation}
where $g_{0}$ is the quadratic-striction parameter. The effect
of defects is taken into account by the term
\begin{equation}
H_{imp} = \int {\rm d^{d}}x\, \left[\Delta{\tau}(x)S^{2}\right]+\int{\rm d^{d}}x\, \left[h(x)\sum\limits_{\alpha} u_{\alpha\alpha}\right],
\end{equation}
where random and Gaussian-distributed variables $\Delta{\tau}(x)$ and $h(x)$ determine local phase-transition temperature
fluctuations and random stress fields, respectively.

To perform calculations, it is convenient to use the
Fourier components of the deformation variables in the
form\begin{equation}
\label{eq:Ftr}
 u_{\alpha\beta} = u^{(0)}_{\alpha \beta} + V^{-1/2} \sum_{q \neq 0}
 u_{\alpha\beta}(q) \exp\left(i q x\right),
\end{equation}
where $q$ is the wavevector, $V$ is the volume, $u^{(0)}_{\alpha \beta}$ is the
uniform deformation tensor, and $u_{\alpha \beta}(q) = {\rm i}/2\left[q_\alpha u_\beta + q_\beta u_\alpha\right]$.
We introduce an expansion in terms of the normal
coordinates,
\begin{equation*}
\vec{u}(q)=\sum_\lambda \vec{e}_\lambda(q) Q_{q,\lambda},
\end{equation*}
where $\vec{e}_\lambda(q)$ is the polarization vector.

We then perform integration with respect to the off-diagonal
components of the uniform part of the deformation
tensor $u^{(0)}_{\alpha \beta}$,
in the statistical sum (they are not
essential for the critical behavior of the system in an
elastically isotropic medium) and obtain a Hamiltonian
for the system in the form of a functional for the spin
order parameter $S\left( q\right)$ and the normal coordinates of the
deformation variables $Q_\lambda\left( q\right)$
\begin{gather}
\label{ham:8}
\tilde{H} = \displaystyle\frac{1}{2}\int {\rm d^{d}}q\ \left(\tau_0+q^{2}\right)\, S_{q}\, S_{-q}
          + \displaystyle\int {\rm d^{d}}q\ q\,h_{q}\, Q_{q,\lambda}  + \nonumber \\
      + a_0\displaystyle\int {\rm d^{d}}q\ q^{2}\, Q_{q,\lambda}\, Q_{-q,\lambda}
          + \displaystyle\frac{1}{2}\int {\rm d^{d}}q\ \Delta{\tau}_{q}\, S_{q_1}\, S_{q_1-q} + \nonumber \\
      + \displaystyle\frac{1}{4}u_0 \int {\rm d^{d}}q\ S_{q_1}\, S_{q_2}\, S_{q_3}\, S_{-q_1-q_2-q_3}- \\
       - \displaystyle w_0 \int {\rm d^{d}}q\ \left(S_{q}\, S_{-q}\right)\, \left(S_{q}\, S_{-q}\right)- \nonumber\\
          -g_0\displaystyle\int {\rm d^{d}}q\ q\, Q_{-q,\lambda}\, S_{q_1}\, S_{q-q_1}.  \nonumber
\end{gather}
where
\begin{align}
w_0&=\displaystyle\frac{3g_0^2}{ 2V\left( 4C_{12}^0 - C_{11}^0 \right)},
&a_0&=\displaystyle\frac{C_{11}^0+4C_{12}^0-4C_{44}}{4V}. \nonumber
\end{align}

The relaxation critical dynamics of compressible
systems is described by dynamic equations of the type
of generalized Langevin equations,
\begin{eqnarray}
\label{dynamic:1}
\dot{S}_{q}=-\Gamma_{0}\,\frac{\partial\tilde{H}}{\partial S_{-q}}+\xi_{q}+\Gamma_{0} h_{S}, \nonumber \\
\label{dinamic:2}
\ddot{Q}_{q,\lambda}=-\frac{\partial\tilde{H}}{\partial Q_{-q,\lambda}} -
q^{2} D_{0} \dot{Q}_{q,\lambda}+\eta_{q}+h_{Q},
\end{eqnarray}
where $\Gamma_{0}$ and $D_{0}$ are bare kinetic coefficients;
$\xi_{q}(x,t)$ and $\eta_{q}(x,t)$ are Gaussian-distributed quantities that
have the character of a random force; and
$h_S$ and $h_Q$ are the fields thermodynamically conjugated to the spin
and deformation variables, respectively.

When solving the set of nonlinear equations (\ref{dynamic:1}) with
Hamiltonian $\tilde{H}(S,Q)$ iteratively, we can single out the
elastic-variable response function $D(q,\omega)$,
which is determined as
\begin{equation}
D(q,\omega) = \displaystyle\frac{\delta \left[\langle{Q_{q,\omega,\lambda}}\rangle\right]}{\delta{h_{Q}}}
= \left[\langle Q_{q,\omega,\lambda}Q_{-q,-\omega,\lambda}\rangle\right],
\end{equation}
and the spin-variable response function $G(q,\omega)$
\begin{equation}
G(q,\omega) = \displaystyle\frac{\delta \left[\langle{S_{q,\omega}}\rangle\right]}{\delta{h_{S}}}
= \left[\langle S_{q,\omega}S_{-q,-\omega}\rangle\right],
\end{equation}
where $\langle ... \rangle$ stands for statistic averaging over random
Langevin forces, $\left[ ... \right]$ stands for the averaging over the
fluctuations of random fields $\Delta{\tau}_{-q}$ and $h_{q}$ that are specified
by structural defects, and $\omega$ is the characteristic
ultrasonic vibration frequency.

Using the Dyson representation, we present the $G(q,\omega)$ and $D(q,\omega)$
response functions in the form
\begin{eqnarray}
\label{DysEq}
G^{-1}(q,\omega) =G_{0}^{-1}(q,\omega) + \Pi(q,\omega), \\
D^{-1}(q,\omega) =D_{0}^{-1}(q,\omega) + \Sigma(q,\omega).
\end{eqnarray}

The bare $G_0(q,\omega)$ and $D_0(q,\omega)$ response functions are
determined as:
\begin{eqnarray}
D_{0}(q,\omega) &=& \left(\omega^{2}-a_0 q^{2}-i\omega D_{0} q^{2}\right)^{-1}, \nonumber\\
G_{0}(q,\omega) &=& \left(i\omega\left/\Gamma_{0}\right.+\left(\tau_0+q^{2}\right)\right)^{-1}. \nonumber
\end{eqnarray}

In the low-temperature phase, the response function
contains an additional relaxation contribution
\begin{equation}
\label{dinamic:1f}
 S_{q}=M\delta_{q,0}+\varphi_{q},
\end{equation}
with the magnetization
\begin{equation}
\label{dinamic:2f}
 M = \left\{
\begin{array}{ll}
0, &  T > T_{c}, \\
B\left|T-T_{c}\right|^{\beta}, &  T < T_{c},
\end{array}
\right.
\end{equation}
where $B$ is the phenomenological relaxation parameter
and $\varphi_{q}$ is the fluctuation part of the order parameter.

The self-energy part $\Sigma(q,\omega)$ of the $D(q,\omega)$ response
function is directly related to the dynamic characteristics
of ultrasound propagation \cite{IroSchwabl}.

\begin{figure}
\includegraphics[width=0.45\textwidth]{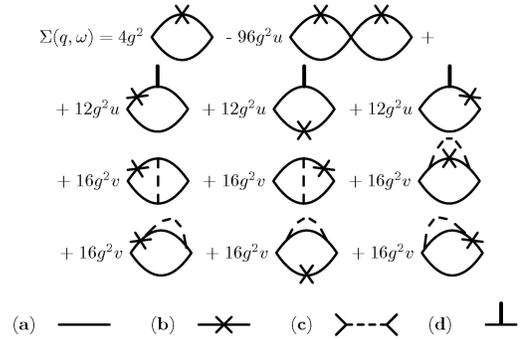}
\caption{ \label{fig:2} Diagrammatic representation of
$\Sigma(q,\omega)$ in a two-loop approximation. Solid line
($\mathbf{a}$) corresponds to $G_{0}(q,\omega)$; lines
($\mathbf{b}$) with a cross, to
$C_{0}(q,\omega)=2\Gamma_{0}^{-1}\left[
\left(\omega/\Gamma_0\right)^2+\left(q^2+\tau_0\right)^2
\right]^{-1}$; vertex with dashed line ($\mathbf{c}$), to
$v=\bigl[(\Delta \tau)^2\bigr]$; and line ($\mathbf{d}$), to
relaxation insertion $M^2\delta_{q,0}$. }
\end{figure}

As a result, the ultrasonic attenuation coefficient can be expressed
through the imaginary part of $\Sigma(q,\omega)$
\begin{equation}
\label{Atten}
    \alpha(\omega,\tau) \sim \omega\,{\rm Im}\Sigma(0,\omega),
\end{equation}
and the sound velocity dispersion is expressed through
its real part,
\begin{equation}
\label{disp}
    c^{2}(\omega,\tau) - c^{2}(0,\tau) \sim \mbox{Re}\left(\Sigma(0,\omega)-\Sigma(0,0)\right).
\end{equation}
We calculated $\Sigma(q,\omega)$ in a two-loop approximation. The
diagrammatic representation of $\Sigma(q,\omega)$ is shown in
Fig.~\ref{fig:2}. These Feynman diagrams contain $d$-dimensional
integration (in our case, $d = 3$).

When approaching the critical point, the correlation length $\xi$
tends to infinity, and, when  $\xi^{-1} \ll \Lambda$ (where
$\Lambda$ is the cutoff parameter of the integration over
wavevectors), the system characteristics demonstrate their
asymptotic scaling behavior for wavevectors $q \ll \Lambda$. Thus,
the calculation of these quantities can be carried out in the limit
$\Lambda \to \infty$. The application of a renormalization-group
procedure eliminates the divergences that appear in the
thermodynamic variables and kinetic coefficients at
 $\Lambda \to \infty$.

To calculate attenuation coefficient (\ref{Atten}) and ultrasound
velocity dispersion (\ref{disp}) and to eliminate the divergences in
 $\Sigma(q,\omega)$ at $q\to~0$ we used the matching
method in \cite{Nelson76}, which was then generalized for the
description of the dynamic behavior of a system in \cite{Matching}.

Thus, using a scaling relationship for the dynamic response function
\begin{equation}
\label{RespF}
D\left(q,\omega,\tau\right) = e^{\left(2-\eta\right) l}D\left(q e^{l},\left(\omega/\Gamma_0\right) e^{zl},\tau e^{l/\nu}\right),
\end{equation}
we can calculate the right-hand side of the equation for some
constant value $l=l^*$, at which not all of the arguments in the
response function disappear simultaneously. The choice of $l^*$ is
determined by the condition
\begin{equation}
\label{MCond}
\left[ \left( \omega/\Gamma_0 \right) e^{zl^*} \right]^{4/z} +
\left[ \left( \tau e^{l^*/\nu} \right)^{2\nu} + q^2 e^{2l^*}\right]^2=1,
\end{equation}
malization- group transformation, namely, to find a relation between
the behavior of the system in the precritical regime at a low value
of reduced temperature $\tau$ and the behavior of the system in a
regime far from the critical mode, i.e., without divergences in
$\Sigma(q,\omega)$. As was demonstrated in \cite{Matching}, matching
condition (\ref{MCond}) provides an infrared cutoff for all
diverging quantities. Based on Eq.~(\ref{MCond}), we find the
solution for $l^*$ in the form of a functional dependence on
$\omega$ and $\tau$, which is specified by the static critical
exponent $\nu$ of the correlation length and by the dynamic critical
exponent $z$,
\begin{equation}
\label{el}
e^{l^*}=\tau^{-\nu} \left[ \,1 +\left(y/2\right)^{4/z} \right]^{-1/4}\equiv \tau^{-\nu}F(y),
\end{equation}
where $y=\omega\tau^{-z \nu}\left/\Gamma_{0}\right.$ is the argument
of the $F(y)$ function.

As is known from the theory of ultrasound scattering in solids near
a phase-transition temperature \cite{Luthi,Suzuki82}, the expression
for the imaginary part of $\Sigma(\omega,\tau)$ in the asymptotic
limit ($\tau \to 0$,\ $\omega \to 0$) can be defined by a scaling
function $\phi(y)$
\begin{eqnarray}
\label{ImScF}
{\rm Im}\Sigma(\omega,\tau) \left/ \omega \right.
&\sim& \tau^{-\alpha-z\,\nu}\phi(y),
\end{eqnarray}
which depends on the single generalized variable $y$. At the same
time, for the imaginary component of the selfenergy part, the
following scaling relationship is valid \cite{IroSchwabl}
\begin{eqnarray} \label{ImSelfR}
& \displaystyle\frac{ {\rm Im}\Sigma(\omega)}{\omega} =
e^{l {\left(\alpha+z\nu\right)}\left/{\nu}\right.}
\displaystyle\frac{ {\rm Im}\Sigma(\omega e^{zl})}{\omega e^{zl}}. \nonumber
\end{eqnarray}
The substitution of $e^{l^*}$ from Eq.~(\ref{el}) into the
right-hand side of this expression allows calculating the $\phi(y)$
scaling function.

In the asymptotic limit $\tau \to 0$,\ $\omega \to 0$), the
expression for the real part of $\Sigma(\omega,\tau)$ can be
determined using another scaling function $f(y)$,
\begin{equation}
\label{ReScF}
\mathrm{Re}\left(\Sigma(0,\omega)-\Sigma(0,0)\right)=\tau^{-\alpha}\left(f(y)-f(0)\right).
\end{equation}
The real component of the self-energy part satisfies the scaling
relation
\begin{multline}
\label{ReSelfR}
\mathrm{Re}\left(\Sigma(0,\omega)-\Sigma(0,0)\right)=\\
=e^{l \alpha\left/{\nu}\right.}\mathrm{Re}\left(\Sigma(0,\omega e^{zl})-\Sigma(0,0)\right).
\end{multline}

The dynamic scaling functions calculated in the two-loop
approximation has the form
\begin{eqnarray}
\label{ScaleFphi} \phi(y) &=&
\displaystyle\frac{g^{*2}\Gamma_0}{\pi\mathstrut}
    \displaystyle\frac{F^{\alpha/\nu+1/2\nu-z}}{y^2}
\left[ 1-  \displaystyle\frac{\left(\Delta+1\right)^{1/2}}{\sqrt{2}}  \right]
- \nonumber \\
 &-& M^{2}
\frac{3g^{* 2}u^*\Gamma_0}{2\pi}
\frac{F^{\alpha/\nu-1/2\nu-z}}{y^{2}}
\left[1-\frac{\left(\Delta+1\right)^{1/2}}{\sqrt{2}\Delta}\right] - \nonumber\\
 &-&\displaystyle\frac{ 3g^{*2}u^*\Gamma_0^2}{\sqrt{2}\pi^2\mathstrut}
    \displaystyle\frac{ F^{\alpha/\nu+1/\nu-2 z}}{ y^3}
         \left[  1- \displaystyle\frac{\left(\Delta+1\right)^{1/2}}{\sqrt{2}} \right]\times \\
 &\times&\left(\Delta -1 \right)^{1/2}-\displaystyle\frac{g^{*2}v^*\Gamma_0}{12\,\pi^3}
        \displaystyle\frac{F^{\alpha/\nu-z}}{y^2}
        \,\ln\,\Delta
, \nonumber\\[3mm]
\label{ScaleFf}
 f(y) &=&
 \displaystyle\frac{g^{*2}\Gamma_0^2}{\pi\mathstrut}
\displaystyle\frac{F^{\alpha/\nu+1/2\nu-z}}{y}
\left[  \displaystyle\frac{\left(\Delta-1\right)^{1/2}}{\sqrt{2}} \right]
- \nonumber \\
&-& M^2\frac{3g^{*2}u^*\Gamma_0^2}{2\pi}\frac{F^{\alpha/\nu-1/2\nu-z}}{ y}
\left[\frac{\left(\Delta-1\right)^{1/2}}{\Delta\sqrt{2}}\right]- \nonumber \\
&-&\displaystyle\frac{ 3\,g^{*2}u^*\Gamma_0^3}{\pi^2\mathstrut}
    \displaystyle\frac{ F^{\alpha/\nu+1/\nu-2 z}}{ y^2}
        \left[  \displaystyle\frac{\left(\Delta+1\right)^{1/2}}{\sqrt{2}} -1 \right]+ \\
&+&
\displaystyle\frac{g^{*2}v^*\Gamma_0^2}{12\pi^3}
        \displaystyle\frac{F^{\alpha/\nu-z}}{y}
        \,\arctan(\Delta^{2}-1)^{1/2},
\nonumber \\[2mm]
 \Delta &=& \left[1+\displaystyle\frac{y^2F^{2z-2/\nu}}{4}\right]^{1/2}, \nonumber
\end{eqnarray}
where $g^*$, $u^*$ and $v^*$ are the magnitudes of the interaction
vertices at the fixed point of the renormalization-group
transformations that corresponds to the critical behavior of the
disordered compressible Ising model \cite{Prudnikov01}. The terms in
Eqs.~(\ref{ScaleFphi}) and (\ref{ScaleFf}) that are proportional to
$M^2$ describe the relaxation contribution for the scaling functions
of the attenuation coefficient and the sound velocity dispersion. In
our subsequent numerical calculations of the scaling functions, we
used the value $\nu=0.70$ from \cite{Prudnikov01} for the
corresponding fixed point. The value of the dynamic exponent
($z=2.1653$) was taken from \cite{Jetp98}, where the critical
dynamics of a disordered Ising model was analyzed within the
framework of a relaxation model. The use of this value of exponent
$z$ is valid in the case of disordered Ising-like systems with a
negative heat-capacity exponent, since the relation between an order
parameter and elastic deformations in the critical dynamics of a
compressible system exerts no substantial influence on the
relaxation properties of the order parameter.

\section{Analysis of results and conclusions}

\begin{figure*}
\includegraphics[width=0.95\textwidth]{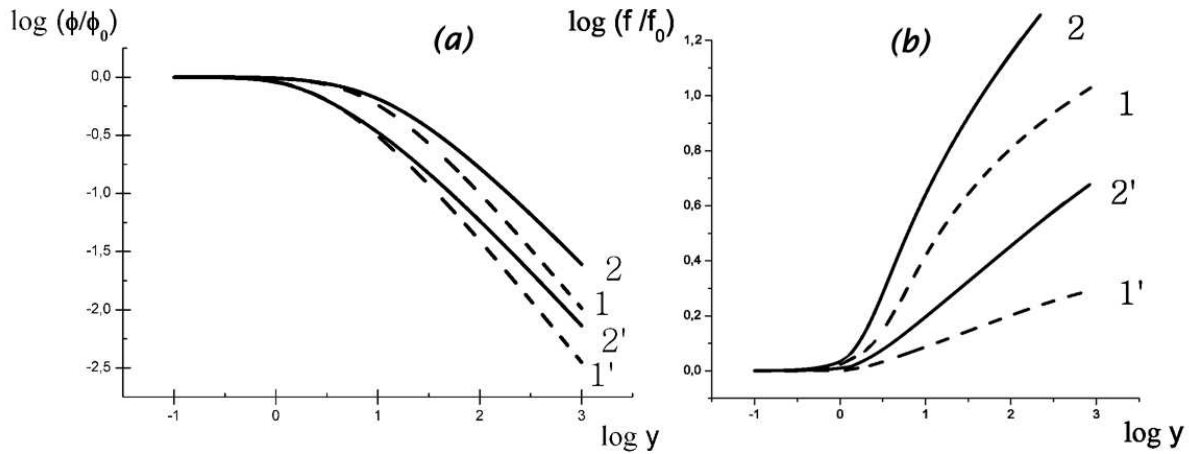}
\caption{ \label{fig:3} Scaling functions (\textbf{a}) $\phi(y)$ and
(\textbf{b}) $f(y)$ for (\textbf{1}) pure and (\textbf{2})
disordered systems at $T>T_c$ and (\textbf{1'}) and (\textbf{2'}) at
$T<T_c$ ($\phi_0=\phi(0), f_0=f(0)$), respectively. }
\end{figure*}

Perturbation-theory series are known to be asymptotic,
and the vertices of the interaction of order-parameter
fluctuations in the fluctuation range
$\tau \rightarrow 0$ are too
high to provide the direct application of Eqs.~(\ref{ScaleFphi}) and (\ref{ScaleFf}).
Therefore, to extract the necessary physical information
from the derived expressions, we apply the
Pad\'{e}–Borel method, which is used to sum up asymptotic
series, that was generalized to a three-parameter
case. Then, the forward and inverse Borel transformations
have the form
\begin{equation}
\begin{array}{rl}
  & \phi(w,u,v)=\sum\limits_{i,j,k}c_{ijk}w^i u^j v^k=\int\limits_{0}^{\infty}e^{-t}B(wt,ut,vt)dt,  \\
  & B(w,u,v)=\sum\limits_{i,j,k}\displaystyle\frac{c_{ijk}}{(i+j+k)!}\,w^i u^j v^k,
\end{array}
\end{equation}
where $w=g^2$.

To analytically continue the Borel transform of the
function, we introduce a series in an auxiliary variable $\lambda$
\begin{equation}
   {\tilde{B}}(w,u,v,\lambda)=\sum\limits_{k=0}^{\infty}\lambda^k\sum\limits_{i=0}^{k}\sum\limits_{j=0}^{k-i}\frac{c_{i,j,k-i-j}}{k!}w^i u^j v^{k-i-j},
\end{equation}
and substitute it to the Pad\'{e} [L/M] approximation at the
point $\lambda=1$.
This procedure was proposed and approved in \cite{Sokolov} to describe the critical behavior of a number of systems
containing several vertices of the interaction of
order-parameter fluctuations. The fact \cite{Sokolov} that the system
retains its symmetry during the application of the
Pad\'{e} approximants in variable $\lambda$ becomes substantial
for the description of multivertex models. In this work,
we calculated the scaling functions in the two-loop
approximation using approximant [1/1].

The behavior of the dynamic $\phi(y)$ and $f(y)$ scaling functions
calculated with summation methods for pure and disordered systems is
shown in Figs.~\ref{fig:3}(a) and~\ref{fig:3}(b) on a log–log scale.
Depending on the interval of changing variable $y$ , the following
asymptotic regions can be distinguished in the behavior of $\phi(y)$
and $f(y)$ : a hydrodynamic region, where $y \sim \omega \xi^z \sim
\left( q \xi \right)^z \ll 1 $, and a critical region $y \sim \omega
\xi^z \gg 1$, which determines the behavior of the system near the
phasetransition temperature ($\tau=(T-T_c)/T_c \ll 1$). As is seen
from these curves at $y\ll 1$, the presence of a structural disorder
does not affect the behavior of the $\phi(y)$ and $f(y)$, scaling
functions and, consequently, the behavior of this system; however,
it begins to manifest itself in the crossover region $10^{-1} < y <
10$ and exerts an essential effect in the critical region $y>10$.

As follows from Eqs.~(\ref{Atten}) and (\ref{ImScF}) , the
attenuation coefficient can be expressed as
\begin{equation}
\label{ass:1}
\alpha\left(\omega,\tau\right)\sim\omega^{2}\tau^{-\alpha-\nu{z}}\phi\left(y\right),
\end{equation}
and, using Eqs.~(\ref{disp}) and (\ref{ReScF}), we can write the
relation for the sound velocity dispersion in the form
\begin{equation}
\label{ass:2}
    c^{2}(\omega,\tau) - c^{2}(0,\tau) \sim \tau^{-\alpha}\left(f(y)-f(0)\right).
\end{equation}

The results of the calculations of the asymptotic dependences of the
attenuation coefficient and the sound velocity dispersion for the
critical and hydrodynamic regions are given in the table. The
characteristics of their frequency and temperature dependences were
determined in the range $10^{-3}\leq y \leq 10^{-1}$, for the
hydrodynamic regime and in the range $10\leq y \leq 10^3$. for the
critical regime. Note that, according to \cite{Matching}, the real
temperature range $10^{-3}\leq \tau \leq 10^{-1}$ in ultrasonic
studies of phase transitions corresponds to the range $1\leq y \leq
10^2$, i.e., it covers the crossover region and the beginning of the
critical region (precritical regime).

\begin{table*}
\caption{Asymptotic behavior of the sound attenuation coefficient
and the sound velocity dispersion in the critical, precritical, and
hydrodynamic regimes for pure and disordered systems}
\begin{ruledtabular}
\begin{tabular}{clcccc}
\multicolumn{6}{c}{Attenuation coefficient $\alpha(\omega,\tau)$}\\
\hline\hline
 \multicolumn{2}{c}{Regim}                      & \multicolumn{2}{c}{Pure}                            & \multicolumn{2}{c}{Disordered }                  \\ \cline{3-6}
                    &                           & $T<T_{c}$                   & $T>T_{c}$                   & $T<T_{c}$                    & $T>T_{c}$                  \\ \hline
 Critical           &  $y=10^{ 1}\div 10^{ 3}$  & $\omega^{0.98}\tau^{-0.08}$ & $\omega^{1.05}\tau^{-0.17}$ & $\omega^{1.12}\tau^{-0.10}$  & $\omega^{1.21}\tau^{-0.24}$\\
 Precritical        &  $y=10^{ 1}\div 10^{ 2}$  & $\omega^{1.08}\tau^{-0.21}$ & $\omega^{1.20}\tau^{-0.37}$ & $\omega^{1.22}\tau^{-0.25}$  & $\omega^{1.37}\tau^{-0.48}$\\
 Hydrodynamic       &  $y=10^{-3}\div 10^{-1}$  & $\omega^{2   }\tau^{-1.38}$ & $\omega^{2   }\tau^{-1.38}$ & $\omega^{2   }\tau^{-1.44}$  & $\omega^{2   }\tau^{-1.44}$\\ \hline\hline
\multicolumn{6}{c}{Sound velocity dispersion
$c^2(\omega,\tau)-c^2(0,\tau)$}\\\hline\hline
 \multicolumn{2}{c}{Regim}                     & \multicolumn{2}{c}{Pure}                            & \multicolumn{2}{c}{Disordered }                  \\ \cline{3-6}
                    &                           & $T<T_{c}$                   & $T>T_{c}$                   & $T<T_{c}$                    & $T>T_{c}$                  \\ \hline
 Critical           &  $y=10^{ 1}\div 10^{ 3}$  & $\omega^{0.11}\tau^{-0.25}$ & $\omega^{0.34}\tau^{-0.54}$ & $\omega^{0.26}\tau^{-0.31}$  & $\omega^{0.49}\tau^{-0.66}$\\
 Precritical        &  $y=10^{ 1}\div 10^{ 2}$  & $\omega^{0.30}\tau^{-0.49}$ & $\omega^{1.08}\tau^{-1.48}$ & $\omega^{0.41}\tau^{-0.54}$  & $\omega^{1.01}\tau^{-1.45}$\\
 Hydrodynamic       &  $y=10^{-3}\div 10^{-1}$  & $\omega^{2   }\tau^{-2.65}$ & $\omega^{2   }\tau^{-2.65}$ & $\omega^{2   }\tau^{-2.95}$  & $\omega^{2   }\tau^{-2.95}$\\
\end{tabular}
\end{ruledtabular}
\end{table*}

It follows from the table that anomalously strong ultrasound
attenuation should be observed in both pure and disordered systems.
For the disordered model, the increase in the attenuation
coefficient as the phase-transition temperature is approached is
expected to be stronger than that in the pure model even in the
hydrodynamic region, whereas in the critical region, the disordered
system should exhibit stronger frequency and temperature dependences
of the attenuation coefficient as compared to the pure system.

\begin{figure}[b]
\includegraphics[width=0.45\textwidth]{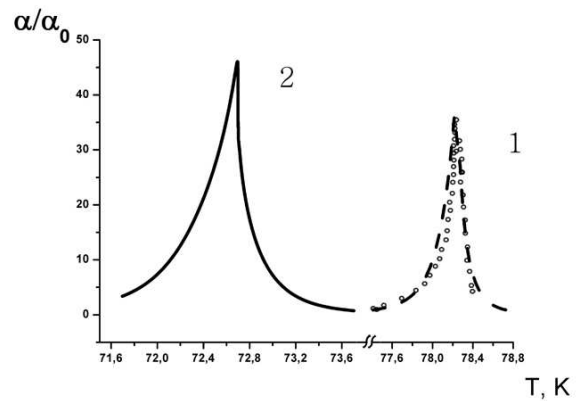}
\caption{ \label{fig:4} Temperature dependence of the attenuation
coefficient calculated for (\textbf{1}) pure and (\textbf{2})
disordered systems at $B=0.3$ and $\omega/\Gamma_0=0.0015$. (dots)
The results of experimental studies of pure
$\mathrm{\mathop{FeF_2}}$ samples \cite{IkushimaF}. }
\end{figure}

These conclusions are supported by the model representation of the
results of the numerical calculations of the critical temperature
behavior of the attenuation coefficient for both the pure and
disordered systems performed at $B=0.3$ and
$\omega/\Gamma_0=0.0015$. (Fig.~\ref{fig:4}). These values were
determined when we compared the calculated temperature dependence of
the attenuation coefficient and the results of experimental studies
of pure $\mathrm{\mathop{FeF_2}}$ samples (Fig.~\ref{fig:4}, (dots)
\cite{IkushimaF},which demonstrate Ising-like behavior in the
critical region.

An analysis of the data related to the sound velocity dispersion
(see table) demonstrates that, as compared to pure analogs, a
structural disorder in Ising-like systems leads a stronger
temperature dependence of the sound velocity dispersion in both the
hydrodynamic and critical regions and is characterized by an
increase in the exponent of the temperature dependence (a decrease
in the absolute value) when going from the hydrodynamic to the
critical region. However, the effect of a structural disorder has
the converse character for the exponent of the frequency dependence
of the sound velocity dispersion: in the hydrodynamic region, the
exponents of these two types of systems coincide, whereas, in the
critical region, the sound velocity dispersion of a disordered
system has a stronger frequency dependence compared to the pure
system. The exponent decreases strongly when going from the
hydrodynamic to the critical region.

A particularly important result of our investigation consists in the
predicted manifestation of the dynamic effects of structural defects
on anomalous sound attenuation and sound velocity dispersion over a
wider temperature range near the critical temperature (already in
the hydrodynamic region) in comparison with other experimental
methods \cite{Rosov}, which require a narrow temperature range (of
about $\tau \simeq 10^{-4}$. to be studied for revealing these
effects. Thus, the results obtained can serve as a reference for
purposeful experimental investigations of the dynamic effects of
structural defects on the critical behavior of solids using acoustic
methods via the detection of the influence of structural defects on
the frequency and temperature dependences of the ultrasound
attenuation coefficient and ultrasound velocity dispersion.



\begin{thebibliography}{99}
\bibitem{IkushimaF}
    A.~Ikushima, R.~Feigelson, J. Phys. Chem. Solids. \textbf{32}, 417 (1971).
\bibitem{Aliev}
    Kh.~K.~Aliev, I.~Kh.~Kamilov, and A.~M.~Omarov, Zh. \'{E}ksp. Teor. Fiz.
    \textbf{95}, 1896 (1989) [Sov. Phys. JETP \textbf{68}, 1096 (1989)].
\bibitem{LandauKh}
    L.~D.~Landau and I.~M.~Khalatnikov, Dokl. Akad. Nauk SSSR \textbf{96}, 496 (1954).
\bibitem{Pawlak}
    A.~Pawlak, Phys. Rev. B \textbf{44}, 5296 (1991).
\bibitem{Schwabl93}
    A.~M.~Schorgg and F.~Schwabl, Phys. Rev. B \textbf{49}, 11682 (1993).
\bibitem{Kamilov98}
    I.~K.~Kamilov and Kh.~K.~Aliev, Usp. Fiz. Nauk \textbf{168},
    953 (1998) [Phys. Usp. \textbf{41}, 865 (1998)].
\bibitem{Bhatt}
    R.~A.~Ferrel, B.~Mirhashem, and B.~Bhattacharjee, Phys. Rev. B \textbf{35}, 4662 (1987).
\bibitem{Luthi}
    T.~J.~Moran and B.~L{\"u}thi, Phys. Rev. B \textbf{4}, 122 (1971).
\bibitem{Suzuki82}
    M.~Suzuki and T.~Komatsubara, J. Phys. C \textbf{15}, 4559 (1982).
\bibitem{Harris74}
    A.~B.~Harris, J. Phys. C \textbf{7}, 1671 (1974).
\bibitem{PawlakFecher89}
    A.~Pawlak and B.~Fechner, Phys. Rev. B \textbf{40}, 9324 (1989).
\bibitem{PrudnikovCM}
    P.~V.~Prudnikov, V.~V.~Prudnikov, J. Phys.: Condens. Matter. \textbf{17}, L485 (2005).
\bibitem{Prudnikov01_03}
    V.~V.~Prudnikov, P.~V.~Prudnikov, A.~A.~Fedorenko, Phys. Rev. B \textbf{62}, 8777 (2000); \textbf{63}, 184201 (2001).
\bibitem{HighOrder}
    R.~Fol'k, Yu.~Golovach, and T.~Yavorskioe, Usp. Fiz. Nauk
    \textbf{173}, 175 (2003) [Phys. Usp. \textbf{46}, 169 (2003)].
\bibitem{Larkin69}
    A.~I.~Larkin and S.~A.~Pikin, Zh. \'{E}ksp. Teor. Fiz. \textbf{56},
    1664 (1969) [Sov. Phys. JETP \textbf{29}, 891 (1969)].
\bibitem{Ymry74}
    Y.~Imry, Phys. Rev. Lett. \textbf{33}, 1304 (1974).
\bibitem{Izym}
    Yu.~A.~Izyumov and V.~N.~Syromyatnikov, Phase Transitions
    and Crystal Symmetry (Nauka, Moscow, 1984; Kluwer, Dordrecht, 1990).
\bibitem{IroSchwabl}
    H.~Iro, F.~Schwabl, Solid State Commun. \textbf{46}, 205 (1983).
\bibitem{Nelson76}
    D.~R.~Nelson,  Phys. Rev. B \textbf{14}, 1123 (1976).
\bibitem{Matching}
    R.~Folk, H.~Iro, F.~Schwabl, Z. Phys. B \textbf{27}, 169 (1977).
\bibitem{Prudnikov01}
    V.~V.~Prudnikov and S.~V.~Belim, Fiz. Tverd. Tela (St.
    Petersburg) 43, 1299 (2001) [Phys. Solid State \textbf{43}, 1353 (2001)].
\bibitem{Jetp98}
    V.~V.~Prudnikov, S.~V.~Belim, A.~V.~Ivanov, et al., Zh.
    \'{E}ksp. Teor. Fiz. \textbf{114}, 972 (1988) [Sov. Phys. JETP \textbf{87}, 527 (1988)].
\bibitem{Sokolov}
    K. B. Varnashev and A. I. Sokolov, Fiz. Tverd. Tela (St.
    Petersburg) \textbf{38}, 3665 (1996) [Phys. Solid State \textbf{38},
    1996 (1996)];
    A.~I.~Sokolov, K.~B.~Varnashev, and A.~I.~Mudrov, Int. J. Mod. Phys. B \textbf{12}, 1365 (1998);
    A.~I.~Sokolov and K.~B.~Varnashev, Phys. Rev. B \textbf{59}, 8363 (1999).
\bibitem{Rosov}
    N.~Rosov, C.~Hohenemser, and M. Eibschutz, Phys. Rev. B \textbf{46}, 3452 (1992).
\end{thebibliography}
\end{document}